# Network Analysis, Plot Theory: Revisiting French Literature through Character Networks

*Newman Chen (nchen1@berkeley.edu), ENS-PSL & CNRS & U. Sorbonne nouvelle, France; Berkeley University, USA*

*Frédérique Mélanie (frederique.melanie@ens.psl.eu), ENS-PSL & CNRS & U. Sorbonne nouvelle, France*

*Jean Barré (jean.barre@ens.psl.eu), ENS-PSL & CNRS & U. Sorbonne nouvelle, France*

*Thierry Poibeau (thierry.poibeau@ens.psl.eu), ENS-PSL & CNRS & U. Sorbonne nouvelle, France*

## 1   Introduction

Character recognition is a technique that enables the automated extraction of characters from texts, while coreference resolution establishes connections between various mentions of the same character, collectively facilitating the creation of expansive character networks (Moretti, 2011). Together, these technologies make it possible to navigate and analyze large literary corpora, opening new avenues for in-depth exploration and understanding of literature. We have created a system specifically for the French language, based on BookNLP-fr (the French counterpart of BookNLP) and NetworkX (a Python package for the manipulation and visualization of complex networks). This allows us to establish connections between series of literary works based on structural features (such as typical relationships between characters) or specific subgenres (for instance, adventure novels featuring a group of young heroes). In this paper, as an illustration, we show the networks obtained at different stages of the short novel *Boule de Suif* from Maupassant (a French 19$^{th}$ century novelist). These figures effectively illustrate how the relationships between the characters develop over the course of the story.

## 2   State of the Art

Significant progress has been made recently in automatically constructing character networks from novels, thanks to advancements in natural language processing and machine learning (Moretti, 2011; Labatut and Bost, 2019). Key methods use advanced tools like Named Entity Recognition (NER) and coreference resolution (Poesio et al., 2023)[1]. The techniques, implemented for example in BookNLP (Bamman et al., 2014, 2020), have proven effective in identifying characters and resolving coreference ambiguities, forming the foundation for character network extraction. As for character networks, representation can be static (i.e. to show the relationships between the characters in a given novel) or dynamic (i.e. to show the evolution of the representation between the main characters in a novel over time) (Labatut and Bost, 2019). Character networks may also help to identify zones in a novel with dense interactions between characters vs zones with fewer characters and fewer interactions (Lavocat, 2020a and 2020b; Rochat, 2014). Interaction can also be defined statistically (i.e. presence of two characters in the same paragraph) or linguistically (i.e. when an explicit syntactic relation expresses a relationship between the two characters), potentially leading to more precise models of relations (who talks to whom, for example).

---

[1] Note that the task has little to do with previous work done for theater, since theater plays directly and explicitly mention a list of characters per scene. Some authors have however proposed some convergence in the techniques, see (Fischer and Skorinkin, 2020).

## 3   Our Approach

In order to extract character network, one thus needs a proper analysis of character mentions, an accurate coreference solver, and then an algorithm to determine when a character interacts with another one. Building upon BookNLP-fr[1] (the French version of BookNLP), our analysis provides a solid basis for identifying characters and resolving coreference ambiguities within French literary works. The model is especially optimized for character recognition, achieving 91.5 precision and 85.2 recall (88.2 F-measure, to be compared to 83.8 when taking into account all entity types, cf. Mélanie et al., 2023). Measures for coreference resolution are more debatable (Duron Tejedor et al. 2023), but the system achieved 77.4 mean F-measure (taking into account Bcub, MUC and CEAF as basic metrics, cf. Mélanie et al., 2024). We think there is still room for improvement here: these models are currently based on a standard version of BERT (Martin et al., 2020), but we plan to move to generative open-source models (e.g.; Llama or Mistral) in the near future, and to fine-tune these models on literature.

Relations between characters are then identified by analyzing cooccurrences at the sentence level, which is appropriate to analyze short novels. This parameter could be changed (so that co-occurrences are based on paragraphs, or on more explicit linguistic markers of interaction between characters, see Grayson et al, 2016), but our approach proved efficient in our context. The corresponding network is finally generated thanks to NetworkX, a python package for "the creation, manipulation, and study of the structure, dynamics, and functions of complex networks" (https://networkx.org/).

## 4   Application to *Boule de Suif* by Maupassant

*Boule de Suif* by Guy de Maupassant is a short story about a prostitute named Elisabeth Rousset, nicknamed "Boule de Suif," who is initially respected for her patriotism but ultimately scorned after she sacrifices her dignity to help her fellow travelers. At the beginning of the story, she is central and indispensable as she earns the group's gratitude by sharing food with them, but by the end, she is rejected and ostracized by the very people she helped, reflecting their hypocrisy and ingratitude. We automatically generated two character networks, representing the relations between characters at the beginning of the short novel, and at the end (**Figure 1 and 2**).

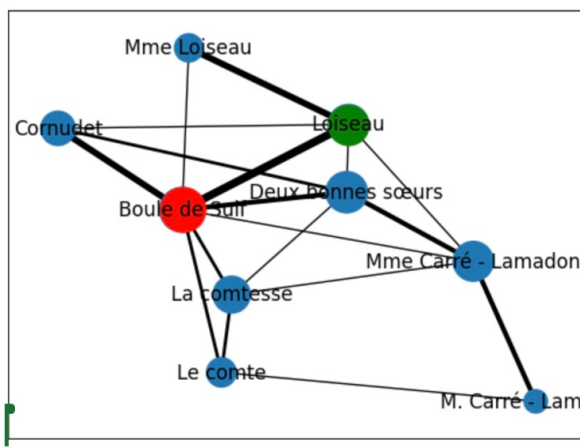 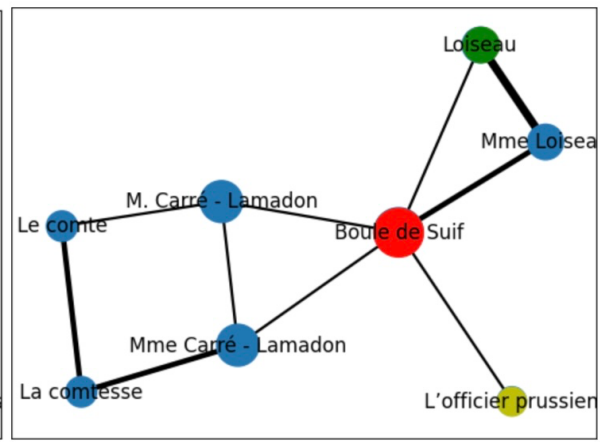

Figure 1                    Figure 2

**Figure 1 and 2.** Character networks representing the relations between characters at the beginning of the short novel *Boule de Suif* **(1)**, and at the end of the novel **(2)**. The color of the nodes reflects their importance in the story.

The graphs clearly show that Boule de Suif is central and highly connected to all the other characters at the beginning, but not so much so in the end. The different characters are also separated, the couple Loiseau being apart from the other ones.

All the characters, independently from their social status, reflect a shared sense of hypocrisy and self-interest. Initially, all the travelers, including Loiseau, Carré-Lamadon, and the Comte and Comtesse, unite in persuading Boule de Suif to sacrifice herself to a Prussian officer for their benefit. Once she does so, they bond over their collective relief and self-righteousness. However, after Boule de Suif complies, they quickly ostracize her, displaying a united front in their ingratitude and moral superiority. Thus, their relationship is one of mutual complicity in exploiting Boule de Suif and then callously rejecting her. In the end of *Boule de Suif*, we also see that the Loiseau couple is not connected to the other characters. The Loiseau couple represents the bourgeoisie, while the other characters (Mr and Mme Carré-Lamadon along with the Comte and Comtesse de Bréville) represent the upper class: they are not really talking to each other due to underlying social and class tensions that resurface after their ordeal, revealing their true feelings and causing them to retreat into their respective social circles, thus highlighting the fragile and superficial nature of their temporary alliance.

Of course, this is an interpretation based on close reading and a good knowledge of the novel, but through this example we aim to show how networks derived from texts can help analyze, interpret, and compare different situations in novels.

## 5 Limitations

Despite the remarkable advancements in character recognition and coreference resolution technologies, the process is not devoid of errors, introducing a layer of complexity in the derivation of character networks from novels. Our character recognition algorithm may occasionally misidentify or miss characters, particularly when faced with ambiguous references or non-standard naming conventions. Similarly, our coreference resolution algorithm can struggle with accurately linking various mentions of the same character, leading to occasional misattribution or omission (Mélanie et al., 2024). These errors, while inherent in the automated nature of these processes, can be propagated to the construction of character networks, potentially distorting the accuracy of relationships and interactions within a narrative.

## 6 Discussion and Conclusion

Character networks provide a visual and structural representation of relationships and interactions within a novel, revealing underlying social dynamics and connections that might be less apparent through traditional textual analysis. By mapping these networks, researchers can gain deeper insights into character roles, thematic elements, and narrative developments, enhancing the overall interpretation and comparison of different literary works. However, close reading remains essential to ensure that interpretations are grounded in a thorough understanding of the book's content.